\documentclass[aps,twocolumn,amsmath,showpacs]{revtex4}
\usepackage{amsmath}
\usepackage{graphicx}

\begin{document}

\title{Nonlinear Saturation of the Weibel Instability in a Dense Fermi Plasma}

\author{F. Haas\footnote{Also at Universidade do Vale do Rio dos Sinos 
- UNISINOS, Av. Unisinos 950, 93022--000, S\~ao Leopoldo, RS, Brazil}, 
P. K. Shukla and B. Eliasson\footnote{Also at the Department of Physics, Ume{\aa} University,
SE-901 87 Ume{\aa}, Sweden}}
\affiliation{Institut f\"ur Theoretische Physik IV, Ruhr--Universit\"at Bochum,
D-44780 Bochum, Germany}
\date{Received 5 March 2008}
\begin{abstract}
We present an investigation for the generation of intense magnetic fields in 
dense plasmas with an anisotropic electron Fermi-Dirac distribution.
For this purpose, we use a new linear dispersion relation for 
transverse waves in the Wigner-Maxwell dense quantum plasma system. 
Numerical analysis of the dispersion relation reveals the scaling of
the growth rate as a function of the Fermi energy and the temperature anisotropy. 
The nonlinear saturation level of the magnetic fields is found through fully kinetic 
simulations, which indicates that the final amplitudes of the magnetic fields are 
proportional to the linear growth rate of the instability.  The present results 
are important for understanding the origin of intense magnetic fields in dense 
Fermionic plasmas, such as those in the next generation intense laser-solid density 
plasma experiments.
\end{abstract}

\maketitle

\section{Introduction}

The existence of feeble magnetic fields of several microgauss in our galaxies \cite{r1}, 
as well as of gigagauss in intense laser-plasma interaction experiments \cite{r2} and 
of billions of gauss in compact astrophysical objects \cite{r3} (e.g. super dense white 
dwarfs, neutron stars/magnetars, degenerate stars, supernovae) is well known. The 
generation mechanisms for seed magnetic fields in cosmic/astrophysical environments 
are still debated, while the spontaneous generation of magnetic fields in laser-produced 
plasmas is attributed to the Biermann battery \cite{r4} (also referred to as the 
baroclinic vector containing non-parallel electron density and electron temperature 
gradients) and to the return electron current from the solid target. Computer simulations 
of laser-fusion plasmas have shown evidence of localized anisotropic electron heating by resonant 
absorption, which in turn can drive a Weibel-like instability resulting in megagauss magnetic 
fields \cite{Estabrook78}. There have also been observations of the Weibel instability in 
high intensity laser-solid interaction experiments \cite{Wei04}. Furthermore, a purely growing 
Weibel instability \cite{r5}, arising from the electron temperature anisotropy 
(a bi-Maxwellian electron distribution function) is also capable of generating magnetic 
fields and associated shocks \cite{r6}. 

However, plasmas in the next generation intense laser-solid density plasma experiments \cite{r8} 
would be very dense. Here the equilibrium electron distribution function may assume the 
form of a deformed Fermi-Dirac distribution due to the electron heating by intense laser beams.
It then turn out that in such dense Fermi plasmas, quantum mechanical effects 
(e.g. the electron tunneling and wave-packet spreading) would play a significant role \cite{r11}. 
The importance of quantum mechanical effects at nanometer scales has been recognized in 
the context of quantum diodes \cite{r12} and ultra-small semiconductor devices \cite{r13}. Also, recently there have been several developments on fermionic quantum plasmas, involving the addition of a dynamical spin force \cite{nr2, nr3, nr4, nr5}, turbulence or coherent structures in degenerate Fermi systems \cite{nr6, nr7}, as well as the coupling between nonlinear Langmuir waves and electron holes in quantum plasmas \cite{d} . The quantum Weibel or filamentational instability for non-degenerate systems has been treated in \cite{nr8, nr9}.   

In this work, we present an investigation of linear and nonlinear aspects of a novel 
instability that is driven by equilibrium  Fermi-Dirac electron temperature anisotropic 
distribution function in a nonrelativistic dense Fermi plasma. Specifically, we show that 
the free energy stored in electron temperature anisotropy is coupled to purely growing 
electromagnetic modes. First, we take the Wigner-Maxwell system \cite{r14} with an anisotropic 
Fermi-Dirac distribution for the analysis of the linearly growing electromagnetic perturbations
as a function of the physical parameters. Second, we use a fully kinetic simulation 
to assess the saturation level of the magnetic fields as a function of the growth rate.
The treatment is restricted to transverse waves, since the latter are associated with the 
largest Weibel instability growth rates.  The nonlinear saturation of the Weibel instability for classical, non-degenerate plasmas has been considered elsewhere \cite{nr1}.

\section{Basic Equations}

It is well known \cite{Pines} that a dense Fermi plasma with isotropic equilibrium distributions 
does not admit any purely growing linear modes. This can be verified, for instance, from the 
expression for the imaginary part of the transverse dielectric function, as derived by 
Lindhard \cite{Lindhard}, for a fully degenerate non-relativistic Fermi plasma. It can be proven 
(see Eq. (30) of \cite{Cockayne}) that the only exception would be for extremely small wavelengths, 
so that $k > 2 k_F$, where $k$ is the wave number and $k_F$ the characteristic Fermi wave number of 
the system. However, in this situation the wave would be super-luminal.  On the other hand, 
in a classical Vlasov-Maxwell plasma containing anisotropic electron distribution function, 
we have a purely growing Weibel instability \cite{r5}, via which dc magnetic fields
are created. The electron temperature anisotropy arises due to the heating of the 
plasma by laser beams \cite{Wei04}, where there is a signature of the  Weibel 
instability as well. In the next generation intense laser-solid density plasma 
experiments, it is likely that the electrons would be degenerate and that  
electron temperature anisotropy may develop due to an anisotropic electron 
heating by intense laser beams via resonant absorption, similar to the classical 
laser plasma case \cite{Estabrook78}.

In a dense laser created plasma, quantum effects must play an important role in the 
context of the Weibel instability. In order to keep the closest analogy with the 
distribution function in phase space for the classical plasma, we shall use the 
Wigner-Maxwell formalism for a dense quantum plasma \cite{Klim}. Here the distribution 
of the electrons is described by the Wigner pseudo-distribution function \cite{Wigner}, 
which is related to the Fermi-Dirac distribution widely used in the random phase 
approximation \cite{Pines}. Proceeding with the time evolution equation for 
the Wigner function (or quantum Vlasov equation \cite{Klim}), we shall
derive a modified dispersion relation accounting for a wave-particle duality
and an anisotropic Wigner distribution function that is appropriate for the Fermi plasma. 
The results are consistent with those of the random phase approximation, in that 
they reproduce the well-known transverse density linear response function for 
a fully degenerate Fermi plasma \cite{Lindhard}. 

Consider linear transverse waves in a dense quantum plasma composed of the  electrons 
and immobile ions, with ${\bf k}\cdot{\bf E} = 0$, where ${\bf k}$ is the wave vector 
and ${\bf E}$ is the wave electric field.  Following the standard procedure, one then 
obtains the general dispersion relation \cite{Klimo,nr9} for the transverse waves 
of the Wigner-Maxwell system  
\begin{eqnarray}
\label{e1}
\omega^2 &-& \omega_{p}^2 - c^2 k^2 + \frac{m \omega_{p}^2}{2n_0 \hbar}
\int d{\bf v}\left(\frac{v_{x}^2 + v_{y}^2}{\omega -
kv_z}\right) \times  \\  &\times& \left(f_{0}(v_{x},v_{y},v_{z} + \frac{\hbar k}{2m}) 
- f_{0}(v_{x},v_{y},v_{z} - \frac{\hbar k}{2m})\right)= 0 \,, \nonumber
\end{eqnarray}
where $\omega$ is the frequency, $c$ is the speed of light in vacuum, $\hbar$ is 
the Planck constant divided by $2\pi$, $m$ the rest electron mass, $n_0$ the 
unperturbed plasma number density, $\omega_p$ the electron plasma frequency, 
${\bf v} = (v_{x}, v_{y}, v_{z})$ is the velocity vector, and $f_{0}(v_{x},v_{y},v_z)$ 
is the equilibrium Wigner function associated to Fermi systems. 

For spin $1/2$ particles, the equilibrium pseudo distribution function is in the form 
of a Fermi-Dirac function. Here we allow for velocity anisotropy and express
\begin{equation}
\label{e2}
f_0 = \frac{\alpha}{\exp\left[\frac{m}{2}\left(\frac{v_{x}^2 + v_{y}^2}{\kappa_{B}T_{\bot}} 
+ \frac{v_{z}^2}{\kappa_{B}T_{\parallel}}\right) - \beta\mu\right] + 1} \,, 
\end{equation}
where $\mu$ is the chemical potential, $\kappa_{B}$ the Boltzmann constant, and 
the normalization constant is 
\begin{equation}
\label{e3}
\alpha = - \frac{n_{0}}{{\rm Li}_{3/2}(- e^{\beta\mu})} \Bigl(\frac{m\beta}{2\pi}\Bigr)^{3/2} 
= 2\Bigl(\frac{m}{2\pi\hbar}\Bigr)^3   \,.
\end{equation}
Here ${\rm Li}_{3/2}$ is a polylogarithm function \cite{r16,r17}. Also, 
$\beta = 1/[\kappa_{B}(T_{\bot}^2 T_{\parallel})^{1/3}]$, where $T_{\bot}$ and $T_\parallel$ are 
related to velocity dispersion in the direction perpendicular and parallel to $z$ axis, respectively. 
In the special case when $T_\bot = T_\parallel$, the usual Fermi-Dirac equilibrium is recovered. 
The chemical potential is obtained by solving the normalization condition (\ref{e3}), yielding, 
in particular, $\mu = {\cal E}_F$ in the limit of zero temperature, where 
${\cal E}_F = (3\pi^2 n_0)^{2/3}\hbar^2/(2m)$ is the Fermi energy. 
Also, the Fermi-Dirac distribution $\hat{f}({\bf k})$, where ${\bf k}$ is the appropriated 
wave vector in momentum space, is related to the equilibrium Wigner function (\ref{e2}) by 
$\hat{f}({\bf k}) = (1/2) (2\pi\hbar/m)^3 f_{0}({\bf v})$, with the factor $2$ coming from 
spin \cite{Ross, Arista}. However, these previous works refer to the cases where there is 
no temperature anisotropy. Notice that it has been suggested \cite{Leemans} that in laser 
plasmas the Weibel instability is responsible for further increase of $T_\parallel$ with time.

Inserting (\ref{e2}) into (\ref{e1}) and integrating over the perpendicular velocity components, we obtain 
\begin{equation}
\label{e4}
\omega^2 - c^2 k^2 - \omega_{p}^2 \left(1 + \frac{T_\bot}{T_{\parallel}}W_Q\right) = 0 \,, 
\end{equation}
where 
\begin{eqnarray}
W_Q &=& \frac{1}{2\sqrt{\pi} H {\rm Li}_{3/2}(-e^{\beta\mu})}\int\frac{d\nu}{\nu-\xi} 
\label{e5} \\ &\times& 
\Biggl({\rm Li}_{2}\Bigl\{-\exp\Bigl[-\Bigl(\nu + \frac{H}{2}\Bigl)^2 + 
\beta\mu\Bigl]\Bigr\} \nonumber 
\\ 
&-& {\rm Li}_{2}\Bigl\{-\exp\Bigl[-\Bigl(\nu - \frac{H}{2}\Bigl)^2 + \beta\mu\Bigl]\Bigr\}\Biggr) \,. \nonumber 
\end{eqnarray}
In (\ref{e5}), ${\rm Li}_{2}$ is the dilogarithm function \cite{r16,r17}, 
$H = \hbar k/(mv_{\parallel})$ is a characteristic parameter representing the 
quantum diffraction effect, $\xi = \omega/(kv_{\parallel})$,  
and $\nu = v_{z}/v_{\parallel}$, with $v_\parallel = (2\kappa_{B}T_{\parallel}/m)^{1/2}$. 
In the simultaneous limit of a small quantum diffraction effect ($H \ll 1$) and 
a dilute system ($e^{\beta\mu} \ll 1$), it can be shown that  $W_Q \simeq - 1 - \xi Z(\xi)$, 
where $Z$ is the standard plasma dispersion function \cite{Fried}. 
It is important to notice that either (\ref{e1}) or (\ref{e4}) reproduces the transverse 
dielectric function calculated from the random phase approximation for a fully degenerate 
quantum plasma \cite{Lindhard}, in the case of an isotropic system. The simple way to verify 
this equivalence is to put $T_\bot = T_{\parallel}$ in (\ref{e1}) and then take the 
limit of zero temperature, so that $f_0 = 3n_{0}/(4\pi v_{F}^3)$ for $|{\bf v}| < v_{F}$, 
and $f_0 = 0$ otherwise, where $v_{F}\equiv (2 {\cal E}_F/m)^{1/2}$ is the Fermi velocity. However, to the best of our knowledge, 
there is no corresponding calculation for an anisotropic Fermi equilibrium, as necessary 
in laser-solid interaction experiments with an anisotropic electron heating due to resonant absorption. 
Also notice that in this Letter we are mainly interested in the real part of the transverse 
response function, since we are looking for purely growing instabilities ($\omega^2 < 0$), 
so that the contribution from the poles at (\ref{e4}) is not relevant. 

\section{Numerical Results}

\begin{figure}[htb]
\centering
\includegraphics[width=8.5cm]{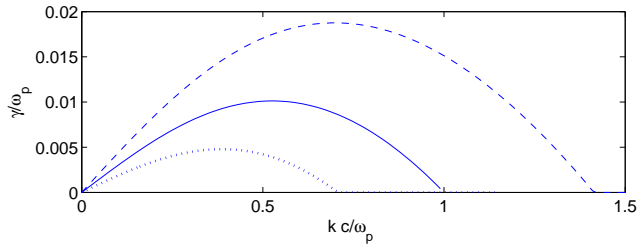}
\caption{The growth rate for the Weibel instability of a dense Fermionic plasma 
with $n_0=10^{33}\,\mathrm{m}^{-3}$ ($\omega_{p}=1.8\times 10^{18}\,\mathrm{s}^{-1}$) and $\beta\mu=5$, 
relevant for the next generation inertially compressed material in intense laser-solid density 
plasma interaction experiments.  The temperature anisotropies are $T_\perp/T_{||}=3$ (dashed line), 
$T_\perp/T_{||}=2$ (solid line) and $T_\perp/T_{||}=1.5$ (dotted line), yielding, respectively, 
$T_{||}=3.9\times 10^6\,\mathrm{K}$, $T_{||}=5.2\times 10^6\,\mathrm{K}$ and $T_{||}=6.3\times 10^6\,\mathrm{K}$. }
\label{Fig1}
\end{figure}

\begin{figure}[htb]
\centering
\includegraphics[width=8.5cm]{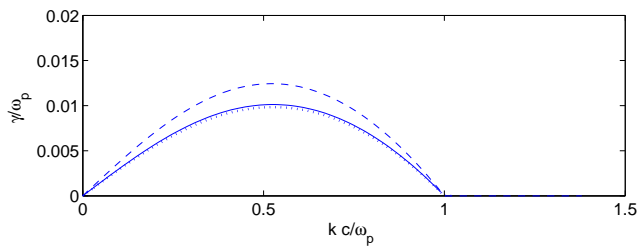}
\caption{The growth rate for the Weibel instability of a dense Fermionic plasma with $n_0=10^{33}\,
\mathrm{m}^{-3}$ ($\omega_{p}=1.8\times 10^{18}\,\mathrm{s}^{-1}$). Here the temperature anisotropy 
is $T_\perp/T_{||}=2$. We used $\beta\mu=1$ (dashed line), $\beta\mu=5$ (solid line) and 
$\beta\mu=10$ (dotted line), yielding $T_{||}=1.6\times 10^7\,\mathrm{K}$, 
$T_{||}=5.2\times 10^7\,\mathrm{K}$ and $T_{||}=2.6\times 10^6\,\mathrm{K}$, respectively.}
\end{figure}

We next solve our new dispersion relation (\ref{e4}) for a set of parameters that are representative 
of the next generation laser-solid density plasma interaction experiments.  The normalization 
condition (\ref{e3}) can also be written as 
$-{\rm Li}_{3/2}[-\exp(\beta\mu)]=(4/3\sqrt{\pi})(\beta{\cal E}_F)^{3/2}$, which 
is formally the same relation holding for isotropic Fermi-Dirac equilibria \cite{Brandsen}. 
For a given value on the product $\beta\mu$ and the density, this relation yields the 
value $\beta$, from which the temperatures $T_\perp$ and $T_{||}$ can be calculated, 
if we know $T_\perp/T_{||}$. Consider only purely growing modes. 
From the definition (\ref{e5}), one can show that $W_Q\rightarrow -1$ when 
$\omega=i\gamma\rightarrow 0$ for a finite wavenumber $k$. 
From (\ref{e4}) we then obtain the maximum wavenumber for instability as 
$k_{\rm max}=(\omega_p/c)\sqrt{T_\perp/T_{||}-1}$. When 
$T_\perp/T_{||}\rightarrow 1$, the range of unstable wavenumbers shrinks to
zero. In Figs. 1 and 2, we have used the
electron number density $n_0=10^{33}\,\mathrm{m}^{-3}$, which can be
obtained in laser-driven compression schemes. The growth rate 
for different values on $T_\perp/T_{||}$ is displayed in Fig. 1. We see
that the maximum unstable wavenumber is $k_{\rm max}=(\omega_p/c)\sqrt{T_\perp/T_{||}-1}$,
as predicted, and that the maximum growth rate occurs at $k\approx k_{\rm max}/2$. 
Figure 1 also reveals that the maximum growth rate of the instability is 
almost linearly proportional to $T_\perp/T_{||}-1$. In Fig. 2, we have varied
the product $\beta\mu$, which is a measure of the degeneracy of the quantum plasma. 
We see that for $\beta\mu$ larger than $5$, the instability reaches a limiting value, 
which is independent of the temperature, while thermal effects start to play an important 
role for $\beta\mu$ of the order unity.

\begin{figure}[htb]
\centering
\includegraphics[width=8.5cm]{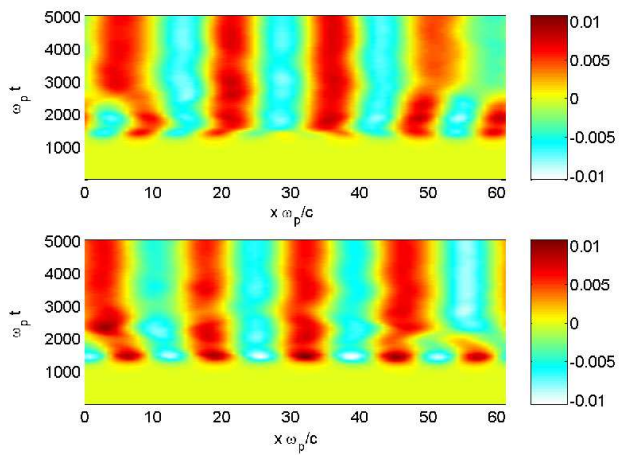}
\caption{The magnetic field components $B_y$ (top panel) and $B_z$ 
(bottom panel) as a function of space and time, for $\beta\mu=5$ and $T_{\perp}/T_{||}=2$.
The magnetic field has been normalized by $\omega_{p} m/e$.  We see a nonlinear saturation 
of the magnetic field components at an amplitude of $\sim 0.01$.}
\end{figure}

\begin{figure}[htb]
\centering
\includegraphics[width=8.5cm]{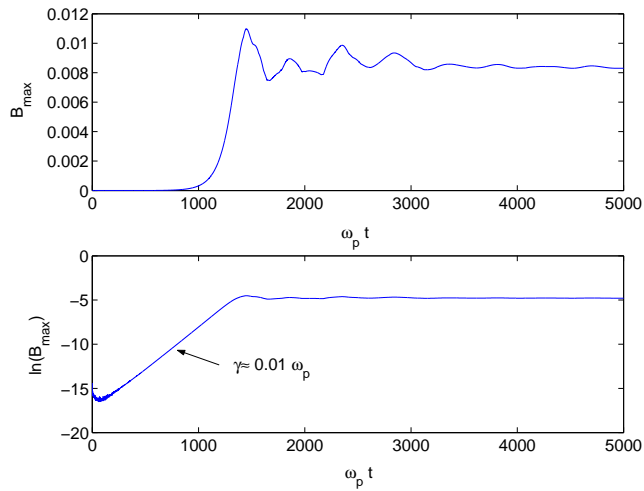}
\caption{The maximum of the magnetic field amplitude, $B=(B_y^2+B_z^2)^{1/2}$, over 
the simulation box (top panel), and the logarithm of the magnetic field maximum (bottom panel) 
as a function of time, for $T_\perp/T_{||}=2$ and $\beta\mu=5$.  The magnetic field has been 
normalized by $\omega_{p} m/e$.  From the logarithmic slope of the magnetic field in the linear regime
we find $\gamma\approx\Delta {\rm ln}(B_{\rm max})/\Delta t\approx 0.01\,\omega_{p}$. }
\end{figure}

From several numerical solutions of the linear dispersion relation, we have been able to 
deduce an approximate scaling law for the instability as 
$\gamma_{max}/\omega_p={\rm constant}\times n_0^{1/3}(T_\perp/T_{||}-1)$,
where the constant is approximately $8.5\times 10^{-14}\,\mathrm{s}^{-1}\mathrm{m}$.
Using that $n_0=(2 m {\cal E}_F/\hbar^2)^{3/2}/(3\pi^2) \approx 1.67\times10^{36}({\cal E}_F/m c^2)^{3/2}$,
we have
\begin{equation}
\label{e7}
  \frac{\gamma_{\rm max}}{\omega_p}=0.10\left(\frac{{\cal E}_F}{m c^2}\right)^{1/2}
  \left(\frac{T_\perp}{T_{||}}-1\right),
\end{equation}
for the maximum growth rate of the Weibel instability in a degenerate Fermi plasma.
This scaling law, where the growth rate depends on the Fermi energy and the temperature anisotropy, 
should be compared to that of a classical plasma \cite{Trivelpiece,Estabrook78}, where the growth rate
depends on the thermal energy and the temperature anisotropy. 

For a Maxwellian plasma, it has been found \cite{Davidson} that the Weibel 
instability saturates nonlinearly once the magnetic bounce frequency 
$\omega_c=eB/m$ has increased to a value comparable to the linear growth rate.  
In order to assess the nonlinear behavior of the Weibel instability for a degenerate plasma, we
have carried out a kinetic simulation of the Wigner-Maxwell system. 
We have assumed that the quantum diffraction effect is small, so that the simulation of the 
Wigner equation can be approximated by simulations of the Vlasov equation by means of an 
electromagnetic Vlasov code \cite{Eliasson07}. As an initial condition for the simulation,
we used the distribution function (2). In order to give a seed for any instability, 
the plasma density was perturbed with low-frequency fluctuations (random numbers).
The results are displayed in Figs. 3 and 4, for the parameters $\beta\mu=5$ and $T_\perp/T_{||}=2$, 
corresponding to the solid lines in Figs. 1 and 2. Figure 3 shows the magnetic field 
components as a function of space and time. We see that the magnetic field initially grows, 
and saturates to steady state magnetic field fluctuations with an amplitude 
of $eB/m\omega_p\approx 0.008$. The maximum amplitude of the magnetic field 
over the simulation box as a function of time is shown in Fig. 4, where we see 
that the magnetic field saturates at $eB/m\omega_p\approx 0.0082$, while the linear 
growth rate of the most unstable mode is $\gamma_{max}/\omega_p\approx 0.009$. 
Similar to the classical Maxwellian plasma case \cite{Davidson},
we can thus estimate the magnetic field (in Tesla) as 
\begin{equation}
  B = \frac{m \gamma_{\rm max}}{e},
\end{equation}
for a degenerate Fermi plasma.  For our parameters parameters relevant for intense laser-solid 
interaction experiments, we will thus have magnetic fields of the order $10^5\,\mathrm{Tesla}$ 
(one gigagauss).

\section{Conclusion}

In conclusion, we have demonstrated the existence of the Weibel instability for 
a Wigner-Maxwell dense quantum plasma, taking into account an anisotropic Fermi-Dirac 
equilibrium distribution function and the quantum diffraction effect. 
Numerically solving the dispersion relation for transverse waves, we found 
the dependence of the growth rate on the Fermi energy and the temperature anisotropy.  
The nonlinear saturation level of the magnetic field was found by means of kinetic simulations, 
which show a linear dependence between the growth rate and the saturated magnetic field. 
The present results may account for intense magnetic fields in dense quantum plasmas, 
such as those in the next generation of intense laser-solid density plasma interaction 
experiments. 

{\bf Acknowledgments}
This work was partially supported by the Alexander von Humboldt Foundation, 
and by the Swedish Research Council (VR).


\begin{thebibliography}{99}
\bibitem{r1}L. W. Wildrow, Rev. Mod. Phys. {\bf 74}, 775 (2002); P. P. Kronberg,
Phys. Today {\bf 55}, 40 (2002).
\bibitem{r2}M. Tatarakis {\it et al.}, Nature (London) {\bf 415}, 280 (2002);
U. Wagner {\it et al.}, Phys. Rev. E {\bf 70}, 026401 (2004).
\bibitem{r3} N. Itoh {\it et al.}, Astrophys. J. {\bf 395}, 622 (1992);
P. Romatschke and R. Venugopalan, Phys. Rev. Lett. {\bf 96}, 062302 (2006).
\bibitem{r4} P. Biermann, Z. Naturforsch. A {\bf 5}, 65 (1950).
\bibitem{Estabrook78} K. Estabrook, Phys. Rev. Lett. {\bf 41}, 1808 (1978).
\bibitem{Wei04} M. S. Wei {\it et al.}, Phys. Rev. E {\bf 70}, 056412 (2004).
\bibitem{r5} E. S. Weibel, Phys. Rev. Lett. {\bf 2}, 83 (1959).
 \bibitem{r6} M. Tzoufras {\it et al.}, Phys. Rev. Lett. {\bf 96}, 105002 (2006).
\bibitem{r8} V. M. Malkin, N. J. Fisch and J. S. Wurtele, Phys. Rev. E {\bf 75}, 026404 (2007).
\bibitem{r11} G. Manfredi, Fields Inst. Commun. {\bf 46}, 263 (2005).
\bibitem{r12} C. L. Gardner and C. Ringhofer, Phys. Rev. E {\bf 53}, 157 (1996).
\bibitem{r13} L. K. Ang and P. K. Zhang, Phys. Rev. Lett. {\bf 98}, 164802 (2007).
\bibitem{nr2} G. Brodin and M. Marklund, New J. Phys. {\bf 9}, 277 (2007)
\bibitem{nr3} M. Marklund, B. Eliasson and P. K. Shukla, Phys. Rev. E {\bf 76}, 067401 (2007).
\bibitem{nr4} G. Brodin and M. Marklund, Phys. Rev. E {\bf 76}, 055403 (2006).
\bibitem{nr5} M. Marklund and G. Brodin, Phys. Rev. Lett. {\bf 96}, 025001 (2007). 
\bibitem{nr6} P. K. Shukla and B. Eliasson, Phys. Rev. Lett. {\bf 96}, 245001 (2006).
\bibitem{nr7} D. Shaikh and P. K. Shukla, Phys. Rev. Lett. {\bf 99}, 125002 (2007).
\bibitem{d} D. Jovanovic and R. Fedele, Phys. Lett. A {\bf 364}, 304 (2007).
\bibitem{nr8} A. Bret, Phys. Plasmas {\bf 14}, 084503 (2007).
\bibitem{nr9} F. Haas, Phys. Plasmas {\bf 15}, 022104 (2008).
\bibitem{r14} F. Haas, Phys. of Plasmas {\bf 12}, 062117 (2005). 
\bibitem{nr1} F. Califano, F. Pegoraro, S. V. Bulanov and A. Mangeney, Phys. Rev. E {\bf 57}, 7048 (1998).
\bibitem{Pines} D. Pines and P. Nozi\`eres, {\it The Theory of Quantum Liquids} (W. A. Benjamin, New York, 1966).
\bibitem{Lindhard} J. Lindhard, K. Dan. Vidensk. Selsk. Mat. Fys. Medd. {\bf 28}, 1 (1954).
\bibitem{Cockayne} E. Cockayne and Z. H. Levine, Phys. Rev. B {\bf 74}, 235107 (2006). 
\bibitem{Klim} Yu. L. Klimontovich and V. P. Silin, in {\it Plasma Physics}, 
edited by J. Drummond (McGraw-Hill, New York, 1961), and many references therein for 
the Wigner-Maxwell method for quantum plasmas. 
\bibitem{Wigner} E. Wigner, Phys. Rev. {\bf 40}, 749 (1932).
\bibitem{Klimo} Yu. L. Klimontovich and V. P. Silin, Zh. Eksp. Teor. Fiz. {\bf 23}, 151 (1952).
\bibitem{r16} M. Abramowitz and I. A. Stegun (eds.), 
{\it Handbook of Mathematical Functions with Formulas, Graphs and 
Mathematical Tables} (Dover, New York, 1972). 
\bibitem{r17} L. Lewin, {\it Po\-ly\-lo\-ga\-rithms and Asso\-cia\-ted Func\-tions} 
(North-Holland, New York, 1981).
\bibitem{Ross} O. Ross, Phys. Rev. {\bf 119}, 1174 (1960).
\bibitem{Arista} N. R. Arista and W. Brandt, Phys. Rev. A {\bf 29}, 1471 (1984). 
\bibitem{Leemans} W. P. Leemans {\it et al.},  Phys. Rev. A {\bf 46}, 1091 (1992).
\bibitem{Fried} B. D. Fried and S. D. Conte, {\it The Plasma Dispersion Function} (Academic Press, London, 1961).
\bibitem{Brandsen} B. H. Brandsen and C. J. Jochain, {\em Introduction to Quantum Mechanics}
(John Wiley \& Sons Inc., New York, 1989).
\bibitem{Trivelpiece} N. A. Krall and A. W. Trivelpiece, {\it Principles of Plasma Physics}
(McGraw-Hill, New York, 1973).
\bibitem{Davidson} R. C. Davidson {\it et al.}, Phys. Fluids {\bf 15}, 317 (1972).
\bibitem{Eliasson07} B. Eliasson, J. Comput. Phys. {\bf 225}, 1508 (2007).
\end{thebibliography}
\end{document}